\documentclass{elsart}
\usepackage[dvips]{epsfig}
\usepackage{citesort}
\newcommand{\kayser}{\ensuremath{\>\mbox{cm}^{-1}}}
\begin{document}
\journal{Chemical Physics Letters}
\begin{frontmatter}
\title{Spin-Lattice Relaxation in Metal-Organic Platinum(II) 
Complexes}
\author{H. H. H. Homeier\thanksref{H4EMA}\thanksref{H4WWW}},
\author{J. Strasser}, and \author{H.
Yersin\thanksref{HY}}
\address{Institut f\"ur Physikalische und Theoretische Chemie,\\
D-93040 Regensburg, Germany}
\thanks[H4EMA]{Author for correspondence. Address: PD Dr. H. H. H. Homeier, 
Institut f\"ur Phy\-si\-ka\-li\-sche und Theo\-re\-ti\-sche Chemie, Universit\"at
Regensburg, D-93040 Regensburg, Germany. FAX: +49-941-943 4719. 
Email: Herbert.Homeier@na-net.ornl.gov}
\thanks[H4WWW]{WWW: http://www.chemie.uni-regensburg.de/$\sim$hoh05008}
\thanks[HY]{Author for correspondence. Address: Prof. Dr. H. Yersin, 
Institut f\"ur  Phy\-si\-ka\-li\-sche und Theoretische Chemie, Universit\"at
Regensburg, D-93040 Regensburg, Germany. FAX: +49-941-943 4488. 
Email: Hartmut.Yersin@chemie.uni-regensburg.de}
\begin{abstract}

The dynamics of spin-lattice relaxation (slr) of metal-organic Pt(II) 
compounds
is studied. Often, such systems are characterized by  
pronounced zero-field splittings (zfs) of the lowest-lying triplets. 
Previous expressions for the Orbach slr process 
do not allow to treat such splitting patterns properly. We discuss
the behavior of a modified Orbach expression for a model system and
present results of a fit of the temperature dependence of
the spin-lattice relaxation rate of 
Pt(2-thpy)$_2$ based on the modified expression. 
\end{abstract}
\end{frontmatter}

Transition metal
complexes with organic chelate ligands 
and their lowest excited states are of potential use for solar 
energy conversion
\cite{r1,r2,r3,r4,r5,r6,r7}. Recently, the processes of spin-lattice relaxation and the 
decay behavior of
excited states  have
been studied experimentally for such systems in Shpol'skii matrices.
 \cite{r9,r12,r13,r15,r16,r17,r18}
Of special importance are 
compounds with a Pt(II) central ion. Pt(II) systems exhibit many different types of low-lying 
excited
triplets that include metal-centered (MC) dd$^{*}$ states
\cite{VanquickenbornCeulemans81,YersinEtal80},
metal-to-ligand-charge-transfer (MLCT) states
\cite{CowmanGray76,YersinGliemann78,GliemannYersin85},
intra-ligand-charge-transfer (ILCT) states
\cite{r13,r17,BallardiniEtal86},
ligand-ligand$'$-charge-transfer (LL$'$CT) states  \cite{r6,r7}, and ligand-centered
(LC) states with some MLCT and/or MC contribution 
\cite{r10,MaestriEtal92,HumbsYersin97}. 
In the following, we focus to
Pt(II) systems with heterocyclic chelate ligands. 

\begin{table}
\caption{Electronic origins E [\kayser] (lowest triplet sublevel
of T$_1$, lowest site), 
zero-field splittings[\kayser] ($\Delta E_{ba}$: 
Energy difference between $\vert b\rangle$ and $\vert a\rangle$,
$\Delta E_{cb}$: Energy difference between $\vert c \rangle$ and $\vert b\rangle$), 
spin-lattice relaxation times $\tau_{\mbox{slr}}$ [ns] at 1.2 K, and transition types  
for various Pt(II)
complexes with organic ligands}\label{tab1}
{\small
\begin{tabular*}{\linewidth}{@{}l@{\extracolsep{\fill}}crrrlr@{}}
\hline\hline
Complex            &  E & $\Delta E_{ba}$ & $\Delta E_{cb}$ & $\tau_{\mbox{slr}}$   & Type       & Ref. \\
\hline
Pt(2-thpy)$_2$$^{a)}$            &           17156 & 7       &  9  &            710 & LC/MLCT    & \cite{r12,r10,r18} \\
Pt(2-thpy)(CO)Cl$^{a)}$          &           18012 & 0.055   &  3.8&           3000 & LC/MLCT    & \cite{r18,GlasbeekHumbsYersin}\\
Pt(phpy)$_2$$^{a)}$              &           19571 & 6.9     & 25.1&            390 & LC/MLCT    & \cite{r18}\\
Pt(3-thpy)$_2$$^{a)}$            &           18020 & 13      &  9  &$\approx$  25   & LC/MLCT    & \cite{r53,Eichenseer99} \\
{}[Pt(bpy)$_2$](ClO$_4)_2$$^{b)}$&           21237 &$<$1     &$<$1 &$>50\cdot 10^3$ & LC/MC      & \cite{HumbsYersin97}\\
Pt(qol)$_2$$^{a)}$               &           15426 &$<$1     &$<$1 &$>60\cdot 10^3$ & ILCT       & \cite{r13,r17} \\
Pt(qtl)$_2$$^{a)}$               &           13158 &$<$1     &$<$1 &        $>$7000 & ILCT       & \cite{r17} \\
Pt(phpy)(CO)Cl$^{a)}$            &           20916 &$<$1     & 6.4 &                & LC/MLCT    & \cite{r52} \\
Pt(bhq)$_2$$^{c)}$               &           19814 & 11      &  28 &                & LC/MLCT    & \cite{BackertYersinZelewsky99}\\
Pt(phpz)$_2$$^{a)}$              &           22952 &  9      &   7 &                & LC/MLCT    & \cite{r53} \\
\hline\hline
\end{tabular*}}\\
2-thpy$^{-}$: 2-(2-thienyl)pyridinate;
phpy$^{-}$: 2,2$'$-phenylpyridinate;
3-thpy$^{-}$: 2-(3-thienyl)pyridinate;
bpy: 2,2$'$-bipyridine;
qol$^{-}$: 8-quinolinolate;
qtl$^{-}$: 8-quinolienthiolate;
bhq$^{-}$: benzo[h]quinolinolate;
phpz$^{-}$: 2,2$'$-phenylpyrazinate.\\ [0.3cm]
$^{a)}$ In n-octane \hskip 2 cm 
$^{b)}$ Neat material\hskip 2 cm
$^{c)}$ In n-decane
\end{table}

As shown in Tab.\ \ref{tab1}, the
low-lying triplets of these systems are characterized by a rather large
variation of
zero-field splittings (zfs)  in the range from less than 0.1 \kayser\ 
to about 40
\kayser. The larger splittings are mainly due to spin-orbit coupling. 
For the same complex in different matrices, the lowest triplet states 
are shifted in energy (in many cases in the range of 200 -- 400
\kayser). The corresponding optical spectra show rich vibrational 
structure
that may be well resolved (about 2\kayser) by choosing appropriate
matrices and by employing methods of 
emission and/or excitation line narrowing.

At low tem\-pe\-rat\-ures (several Kelvin),
the processes of spin-lattice relaxation occurring between the 
triplet sublevels
$\vert a\rangle$, $\vert b\rangle$, and $\vert c\rangle$ are
relatively slow with relaxation times
as long as hundreds of nano-seconds and even up to many micro-seconds 
(See Tab.\ \ref{tab1} and Refs.\ 
\cite{r12,r13,r15,r16,r17,r18}) due to the low density 
of phonon states corresponding
to such zfs patterns.

To discuss these processes, we assume that the 
perturbation $V$ caused by the
phonons couples the electronic states
of the chromophore essentially linearly (e.g.\ see Ref. 
\cite[p. 228]{r38})
\begin{equation}
V = V_1 \sum_{k}\epsilon_k + \dots
\end{equation}
where $\epsilon_k$ is the strain corresponding to the phonon mode with
wave vector $k$ in the long wavelength limit. The matrix elements of
$V_1$ are denoted by $V_{{ b}{ a}}=\vert \langle { b} \vert V_1 \vert
{ a}\rangle \vert$ and analogous expressions for $V_{{ c}{ a}}$ and
$V_{{ c}{ b}}$. 
The energy differences are $\Delta E_{ b a}$ between $\vert  b\rangle$ 
and $\vert  a\rangle$,
$\Delta E_{ c b}$ between $\vert  c\rangle$ and $\vert  b\rangle$, and
$\Delta E_{ c a}$ between $\vert  c\rangle$ and $\vert  a\rangle$.
The usual notation $\beta=1/(k_B T)$ for given
tem\-pe\-rat\-ure $T$ and
Boltzmann constant $k_B$, and the abbreviations $ C_{{ b}{ a}} =  C\,
V_{ b a}^{2} \left({\Delta E_{{ b}{ a}}}\right)^{3} $ and analogous
ones for $C_{{ c}{ a}}$ and $C_{{ c}{ b}}$ are also used. Here, the
parameter $C={3}/({2\pi\hbar^4\rho v^5})$ is defined in terms of mass
density  $\rho$ and  (average) velocity $v$ of sound of the matrix. 
The $(\Delta E_{ b a})^3$ dependence of $C_{ b a}$ should be kept 
in mind.

\begin{figure}
\begin{center}
\epsfig{file=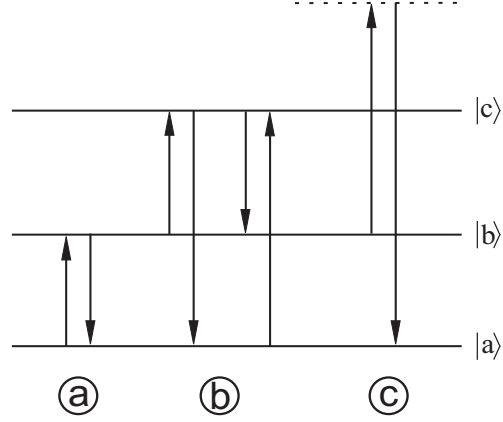,width=6.5cm,height=5.5cm}
\end{center}
\caption{
Processes of spin-lattice relaxation: 
a)~Direct process. b)~Orbach process. c)~Raman process.}\label{fig2}
\end{figure}

The following relaxation processes (see Fig.\ \ref{fig2})
occur:

\paragraph*{Direct process:} The rate is given by
\cite[p. 541]{r37},
\cite[p. 229]{r38}
\begin{eqnarray}\label{eqdirect}
k^{(direct)}_{{ a},{ b}}&=& k_{ a b}+k_{ b a} 
   \nonumber\\
                        &=& C_{ b a}\,
                            \coth(\beta \Delta E_{{ b}{ a}}/2)\>.
\end{eqnarray}
Here, $k_{{ a}{ b}}$ and $k_{{ b}{ a}}$ are the rate constants for the 
up and
down processes, respectively, given by the expressions
\begin{eqnarray}\label{solution}
k_{{ a}{ b}}&=&C_{ b a}
               \frac{1}
               {\exp(\beta\Delta E_{{ b}{ a}})-1} \>, 
   \nonumber \\
k_{{ b}{ a}}&=&C_{ b a}
               \frac{\exp(\beta\Delta E_{{ b}{ a}})}
                    {\exp(\beta\Delta E_{{ b}{ a}})-1}\>.
\end{eqnarray}
Analogous expressions hold for the up and down rates 
$k_{ b c}$, $k_{ c b}$,
$k_{ a c}$, and $k_{ c a}$.

\paragraph*{Orbach process:} The rate for this process
vanishes for $T\to 0$~K exponentially. It
depends on the splitting pattern of the three involved states: 
If the energy
separation $\Delta E_{ b a}$ of the two lower
states $\vert  a\rangle$ and $\vert  b\rangle$ is much smaller than both
the energy separations $\Delta E_{ c a}$ and $\Delta E_{ c b}$ to the
upper state $\vert  c\rangle$, then the well-known expression
\begin{equation}\label{eqorgOrbach}
k^{(Orbach)}_{{ a},{ b}} =
 \frac{2 C_{{ c}{ b}} C_{{ c}{ a}}
}%
{(C_{{ c}{ a}}
+C_{{ c}{ b}})} \> \frac{1}{\displaystyle(\e^{\beta \Delta E}-1)}\>
\end{equation}
holds approximately for low $T$. This original Orbach expression is 
derived under the
assumption that the energy differences are
given by
$\Delta E=\Delta E_{{ c}{ a}}=\Delta E_{{ c}{ b}}>0$.
For a more general zfs pattern, the rate is given by
the low-tem\-pe\-rat\-ure approximation \cite{StrasserHomeierYersin99}
\begin{equation}
\label{eqnewOrbach}
k^{(Orbach)}_{{ a},{ b}} =
\frac{k_{{ a}{ c}} k_{{ c}{ b}}+k_{{ b}{ c}} k_{{ c}{ a}}-k_{{ b}{ c}} 
k_{{ b}{ a}}}{k_{{ c}{ a}}+k_{{ c}{ b}}-k_{{ b}{ a}}}
\end{equation}
with up and down rates as given in Eq.\ (\ref{solution}). 
The modified expression 
(\ref{eqnewOrbach})
contains Eq.\ (\ref{eqorgOrbach}) as a limiting case 
(see Ref.\ \cite{StrasserHomeierYersin99}).

\paragraph*{Raman process:} For low tem\-pe\-rat\-ure, the rate may  be
approximated by
\begin{equation}\label{eqRaman}
k^{(Raman)}_{{ a},{ b}} = D\, T^n
\end{equation}
with a constant $D$ and $n=5$ for non-Kramers ions \cite{r39a}. In the
cases under study, this $T^5$ dependence fits the experimental
observations \cite{StrasserHomeierYersin99} better than the $T^7$ dependence observed in other
systems. 

The relative importance of the various slr processes is 
largely dependent on the size of
the zfs and the energy separations to  further electronic states. 
For instance, in systems like Pt(qol)$_2$ and  Pt(qtl)$_2$ 
with a very small total zfs (see Tab.\ \ref{tab1}) and 
no further electronic states in the  vicinity of T$_1$, 
direct and Orbach processes are expected to be very small
due to the $\Delta E^3$ dependence of these processes, and 
the Raman process is expected to dominate. 
Compare also Ref.\ \cite{StrasserHomeierYersin99}.

\begin{figure}
\begin{center}
\begin{minipage}{9cm}
{\epsfig{file=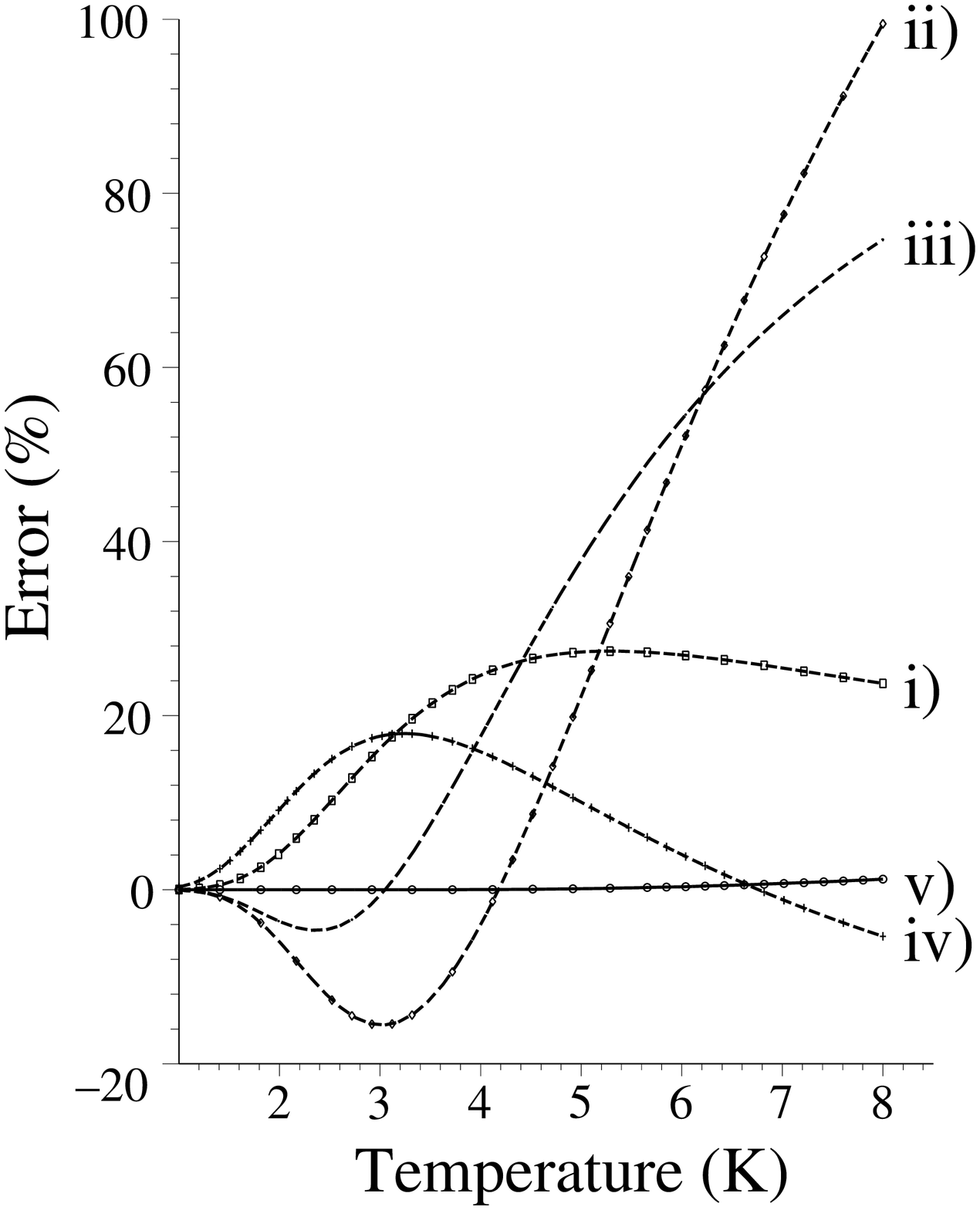,width=9 cm,height=10 cm}}

\vskip -10 cm
\hglue 2.5 cm\epsfig{file=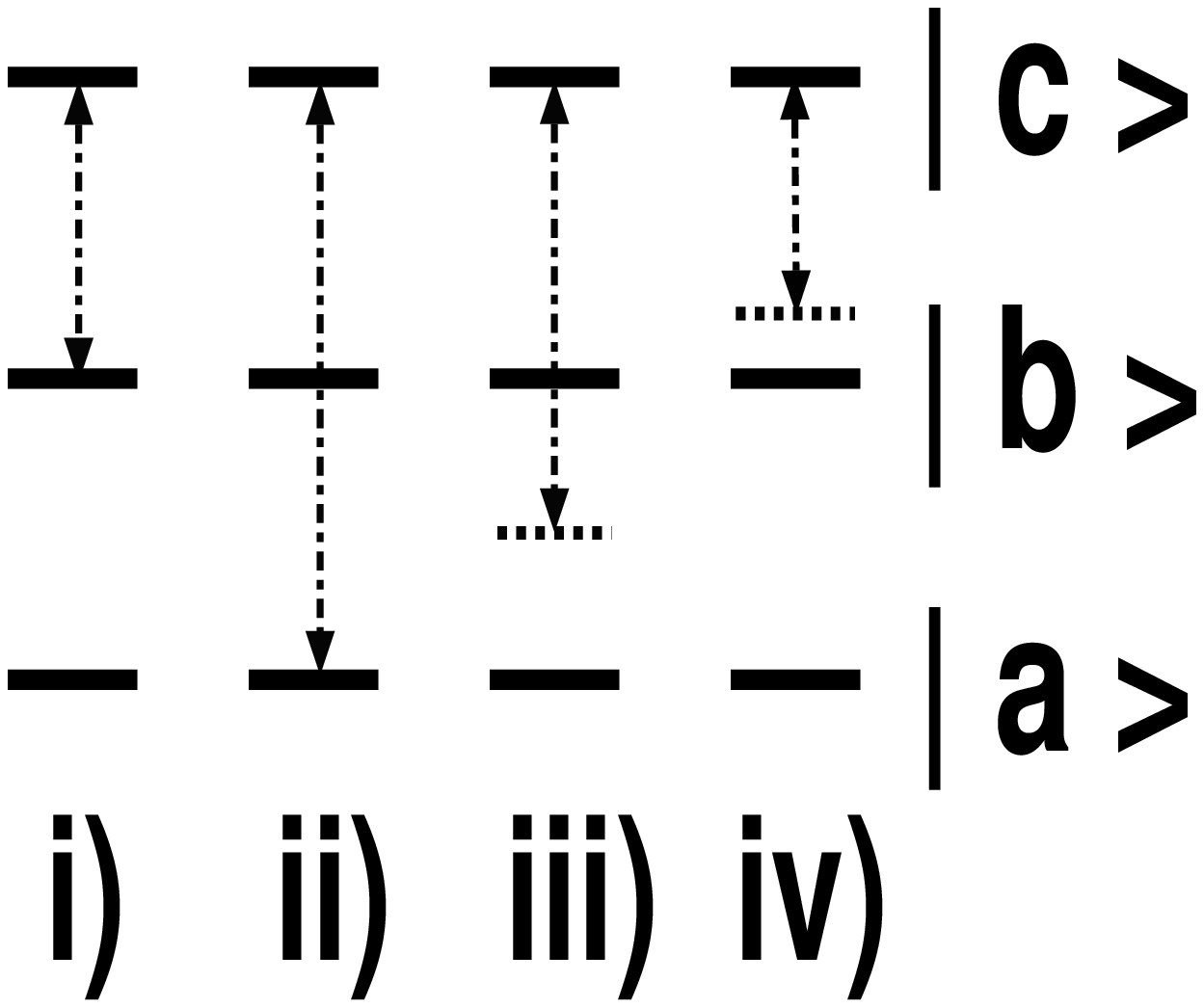,width=3.5cm,height=3 cm}
\end{minipage}

\vskip 7 cm
\end{center}
\caption{
Relative errors of the relaxation rate expressions with respect to Eq.\ (\ref{exactdrei}) 
as a function of
tem\-pe\-rat\-ure $T$. Plotted are the errors of
$k_{ a, b}^{(direct)}+k_{ a, b}^{(Orbach)}$ with  Eq.\
(\ref{eqdirect}) for the direct process in combination with the 
original Orbach expression (\ref{eqorgOrbach}) for different values 
of $\Delta
E$ ( i)
$\Delta E=\Delta E_{ c b}= 7\kayser$, ii)
$\Delta E=\Delta E_{ c a}=14 \kayser$,  iii)
$\Delta E=(\Delta
E_{ c a}+\Delta E_{ c b})/2=10.5\kayser$,  iv)
$\Delta E=\Delta E_{\mbox{fit}}=
5.4 \kayser$
) and v) with the modified expression (\ref{eqnewOrbach}).
}\label{figsimulhy}
\end{figure}

The behavior of the above  expressions is illustrated for a \emph{model}
system (without a Raman process) and with parameters 
$\Delta E_{ b a}=\Delta E_{ c b}=7 \kayser$,
$v=2000$ m/s, $\rho=1.1$ g/cm$^{3}$,
$V_{ b c}=10 \kayser$, $V_{ a c}=20 \kayser$, $V_{ a b}=3 \kayser$. 
In
Fig.\ \ref{figsimulhy}, the relative errors of the approximations 
for both direct and Orbach process, i.e., for the sum
$k_{ a, b}^{(direct)}+k_{ a, b}^{(Orbach)}$ as obtained using Eq.\
(\ref{eqdirect}) in combination either with Eq.\ (\ref{eqorgOrbach}) or
Eq. (\ref{eqnewOrbach}), respectively. The errors are calculated with 
respect to the exact
rate
\begin{eqnarray}\label{exactdrei}
\lefteqn{k^{(Orbach+direct)}_{{ a},{ b}} } \nonumber\\
&=&\frac{1}{2} \biggl(
    k_{{ b}{ c}}+k_{{ a}{ c}}+k_{{ c}{ b}}
   +k_{{ c}{ a}}+k_{{ b}{ a}}+k_{{ a}{ b}}\biggr) \nonumber\\
&-&\frac{1}{2}\biggl((k_{{ b}{ c}}+k_{{ c}{ b}}-k_{{ a}{ b}}
           -k_{{ c}{ a}}-k_{{ a}{ c}}+k_{{ b}{ a}})^2 \nonumber\\
&+&4(k_{{ c}{ b}}k_{{ c}{ a}}-k_{{ a}{ b}}k_{{ c}{ a}}
    - k_{{ b}{ a}}k_{{ c}{ b}}+k_{{ b}{ a}}k_{{ a}{ b}})
    \biggr)^{1/2}
\end{eqnarray}
for the three-level system 
that is obtained from the rate equations \cite{StrasserHomeierYersin99}.

Applying the original Orbach expression, i.e.,
using Eq.\  (\ref{eqorgOrbach})  in combination with (\ref{eqdirect}) 
for the direct process, 
the prefactor $2 C_{{ c}{ b}} C_{{ c}{ a}} /{(C_{{ c}{ a}}
+C_{{ c}{ b}})}$ was computed from the model parameters, but different 
values of
the parameter $\Delta E$ have been used: $\Delta E=\Delta E_{ c b}$
corresponds to using the minimum distance of state $\vert c\rangle$ to
the states $\vert  a\rangle$ and $\vert  b\rangle$ (curve i) in Fig.
\ref{figsimulhy}), $\Delta E=\Delta E_{ c a}$ corresponds to using the
maximum distance (curve ii)), and $\Delta
E=(\Delta E_{ c a}+\Delta E_{ c b})/2$ corresponds to using the mean
distance (curve iii)). The value $\Delta
E=\Delta E_{\mbox{fit}}= 5.4 \kayser$ is obtained by a least square fit
of the exact data with one fit parameter $\Delta E$ (curve iv)), 
i.e., for the direct process and the prefactor of Eq.\  
(\ref{eqorgOrbach}), 
the exact expressions were used during the fit.
 Interestingly, $\Delta E_{\mbox{fit}}$ is less than
any of the other differences of the energies. Alternatively, one 
could try to use
the prefactor  in  Eq.\ (\ref{eqorgOrbach}) as an additional fit
parameter. But then, one cannot hope to extract the model values of
$C_{ c b}$ and $C_{ c a}$ from such a fit. Finally, curve v) in Fig.
\ref{figsimulhy} was obtained using the modified expression 
(\ref{eqnewOrbach}) in combination with Eq.\ (\ref{eqdirect}) for 
the direct process. 
Clearly,
the modified approach yields much reduced errors over a large
tem\-pe\-rat\-ure range. Thus, Orbach's original expression 
(\ref{eqorgOrbach}) that was
designed for a different pattern of the energy levels cannot be 
applied to a pattern
with $\Delta E_{cb}\approx \Delta E_{ba}$ for any reasonable choice of
the parameter $\Delta E$.

We remark that similar results are also obtained for different 
choices of the parameters. 
For instance, for a value of $v$ smaller by a factor $f$,
the same results for the absolute rates would be obtained, if all the 
matrix elements of 
$V_1$ are also chosen smaller by a 
factor $f^{5/2}$, e.g., for
$v=1500$ m/s and $V_{ b c}=4.87 \kayser$, $V_{ a c}=9.74 \kayser$, 
$V_{ a b}=1.46 \kayser$. 
Moreover, fixing all the other parameters, 
any rescaling of the 
three matrix elements by an arbitrary common positive factor 
yields the same  error curves since we are dealing with \emph{relative}
errors and, under this scaling, 
all up and down rates $k_{ a b}$,
$k_{ b a}$ etc., and, hence, all slr rates in the model 
are multiplied by a common factor.

\begin{figure}
\begin{center}
{\epsfig{file=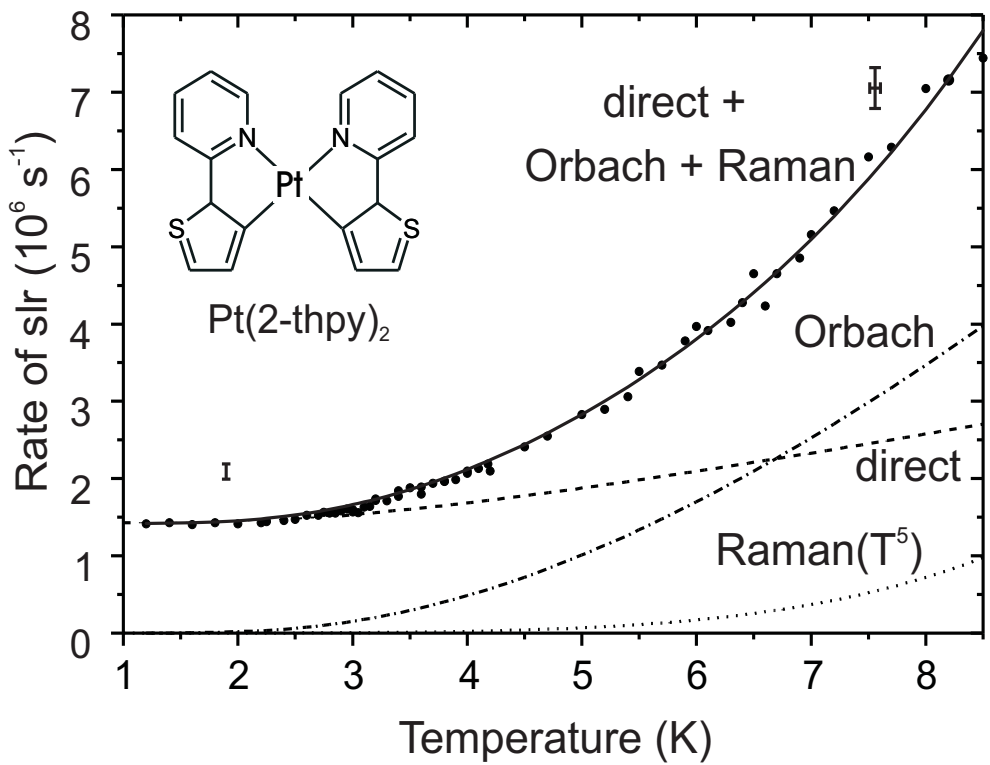,width=13 cm,height=10 cm}}
\end{center}
\caption{Fit of the spin-lattice relaxation rate 
$k^{(slr)}$ as a function of tem\-pe\-rat\-ure for
Pt(2-thpy)$_2$ in an n-octane matrix. Displayed are the
contributions of the direct process (Eq.\ (\ref{eqdirect})), 
the Orbach process (using the modified expression
(\ref{eqnewOrbach})), and the Raman $T^5$ process (Eq.\
(\ref{eqRaman})).}\label{fig3}
\end{figure}

It is of interest to present an example of the application of the 
above formalism 
to the spin-lattice relaxation observed for the lowest triplet of
the Pt(2-thpy)$_2$ 
complex in an n-octane matrix. This compound is
depicted in of Fig.\ \ref{fig3}, and some properties are
collected in Table \ref{tab1}. The experimental spin-lattice
relaxation rate $k^{(slr)}$ is obtained from the measured 
emission decay rate of state
$\vert b\rangle$  
by subtraction of the corresponding 
triplet deactivation rate to the ground state 
\cite{StrasserHomeierYersin99}. For the fit, we used 
Eq.\ (\ref{eqdirect}) for the direct process, 
the modified expression
(\ref{eqnewOrbach}) for the Orbach process, and Eq.\
(\ref{eqRaman}) with $n=5$ for the Raman process, i.e., for a $T^5$
low tem\-pe\-rat\-ure dependence. 
As prefactor of the direct process, we used the low 
tem\-pe\-rat\-ure limit
of $k^{(slr)}$. The ratio of $C_{ c a}/C_{ c b}$ can be obtained
independently from time-resolved excitation spectra
\cite{r12,StrasserHomeierYersin99}. Also, all energy separations
$\Delta E_{ b a}$ and $\Delta E_{ c b}$ are available from highly
resolved spectra \cite{r12,r10,r18,StrasserHomeierYersin99}. 
Thus, as fit parameter, only the prefactor $D$
of the Raman process and the constant $C_{ c a}$ remain. 
For such a two-parameter fit as
displayed in Fig.\ \ref{fig3},
the result is highly satisfactory. 

A three-parameter fit 
based on the original 
Orbach expression (\ref{eqorgOrbach}) using the parameters 
$D$, $\Delta E$ and the
prefactor in Eq.\ (\ref{eqorgOrbach}) yields the  
value $\Delta E=11.4 \kayser$ 
(and a nearly doubled prefactor $D$ for the Raman process in 
comparison to the fit displayed in   
Fig.\ \ref{fig3}). A similar value for $\Delta E$ was obtained 
in Ref.\ \cite{r12} by
a somewhat different fitting procedure. Both these values are
unphysical since they do not correspond to any of the observed 
energy differences (see Tab.\ \ref{tab1}). We remark that 
the present study was  triggered by this difficulty of using the 
original Orbach expression (\ref{eqorgOrbach}).

This result shows, as further ones  presented in 
\cite{r58,StrasserHomeierYersin99}, that
the use of the modified expression (\ref{eqnewOrbach}) 
for the Orbach process is necessary for a
detailed understanding of  the dynamics of the spin-lattice 
relaxation for low-lying triplets of
metal-organic transition metal compounds with their 
characteristic patterns of zero-field splitting. Thus, although
the present study concentrated on Pt(II) compounds, 
the result should be applicable to
a more general class of compounds, namely, 
the whole platinum metal group complexes (compare, e.g., the recent
results \cite{r9,r12} for [Ru(bpy)$_3$]$^{2+}$).

Financial support by the \emph{Deutsche Forschungsgemeinschaft} and the
\emph{Fonds der Chemischen Industrie} is
gratefully acknowledged.

\vfill
\break

\vfill

\end{document}